\documentclass[preprint,aps,showpacs,amsmath,amssymb,prd,eqsecnum]{revtex4-1}
\usepackage{color}
\usepackage{natbib}
\allowdisplaybreaks

\begin{document}
\bibliographystyle{revtex}

\title{Quantization of the minimal and non-minimal vector field in curved space}

\author{David J. Toms}
\homepage{http://www.staff.ncl.ac.uk/d.j.toms}
\email{david.toms@newcastle.ac.uk}
\affiliation{
School of Mathematics and Statistics,
Newcastle University,
Newcastle upon Tyne, U.K. NE1 7RU}

\date{\today}

\begin{abstract}
The local momentum space method is used to study the quantized massive vector field (the Proca field) with the possible addition of non-minimal terms. Heat kernel coefficients are calculated and used to evaluate the divergent part of the one-loop effective action. It is shown that the naive expression for the effective action that one would write down based on the minimal coupling case needs modification. We adopt a Faddeev-Jackiw method of quantization and consider the case of an ultrastatic spacetime for simplicity. The operator that arises for non-minimal coupling to the curvature is shown to be non-minimal in the sense of Barvinsky and Vilkovisky. It is shown that when a general non-minimal term is added to the theory the result is not renormalizable with the addition of a local Lagrangian counterterm.  
\end{abstract}


\maketitle

\section{Introduction}\label{sec-intro}

The currently accepted mechanism for giving masses to the vector fields in the standard model is by the Higg's mechanism~\cite{higgs1964broken1,higgs1964broken,englert1964broken,guralnik1964global}. Nevertheless there has been continuing interest in the simple addition of mass terms to the standard vector Lagrangian as originated in the work of Proca~\cite{proca1936theorie}. See also \citep{pauli1941relativistic} and \citep{Wentzel} for early reviews. An alternative approach, due to Stueckelberg, is reviewed in \citep{ruegg2004stueckelberg}.

The main goal of the present paper is to examine the massive quantized vector field (the Proca field) in curved spacetime, particularly in the presence of possible non-minimal terms of the form assumed in \eqref{Lagrangian} with \eqref{nonmin} below. Previous work on the quantized massive vector (without the non-minimal terms) includes Furlani~\citep{furlani1997quantization} who looked at canonical quantization in an ultrastatic spacetime, Barvinsky and Vilkovisky~\citep{BarvinskyVilkovisky} who looked at the effective action on a general background, Gorbar and Shapiro~\citep{gorbar2003renormalization} who performed a renormalization group analysis, and Buchbinder, de Berredo-Peixoto, and Shapiro~\citep{buchbinder2007quantum} who used a Stueckelberg analysis. More recently Prescod-Weinstein and Bertschinger~\citep{prescod2015extension} examined a more complicated Landau-Ginzberg type theory in curved spacetime which includes a massive vector field. They also utilized the Faddeev-Jackiw~\citep{FaddeevJackiw} method that we will say more about later. The non-minimal terms that we will consider were apparently first introduced by Novello and Salim~\citep{novello1979nonlinear} who looked at their effect on cosmology but at the classical level. The first study in quantum theory was in \citep{davies1985boundary} where it was shown how such terms might possibly lead to an observational effect in a Casimir type calculation.

In the next section we will consider how the local momentum space approach introduced by Bunch and Parker~\citep{BunchParker} may be used to study the heat kernel coefficients of the differential operator that arises in the quantized massive vector field. This presents an alternative to the approach of Barvinsky and Vilkovisky~\citep{BarvinskyVilkovisky} that is perhaps more straightforward, but also more tedious. The results of our earlier work~\citep{toms2014local} are used and we calculate the first few heat kernel coefficients and the divergent part of the one-loop effective action. We point out a problem in the evaluation of terms in the one-loop effective action, or equivalently in the relevant heat kernel coefficient, that are total derivatives. In Sec.~\ref{sec_Procaquantized} we examine in detail the correct quantization of the non-minimally coupled Proca field using the Faddeev-Jackiw method~\citep{FaddeevJackiw}. We give a reason from the viewpoint of constrained quantization why the identity used in Barvinsky and Vilkovisky~\citep{BarvinskyVilkovisky} (see \eqref{BVidentity} below) works in the minimal case. Based on this analysis we calculate the divergent part of the one-loop effective action for the minimal Proca field again in an ultrastatic spacetime and show that the result agrees with our earlier one and with \citep{BarvinskyVilkovisky}; this second method, unlike the first, does not require the usual $\delta(0)\rightarrow0$ regularization of dimensional regularization. In Sec.~\ref{sec-flat} we examine the renormalizability of the Proca field where a non-minimal term is added. It is shown that it is not possible to renormalize the theory by adding a local counterterm to the Lagrangian. We also discuss why the heat kernel method results in a misleading divergence in this case.

\section{Proca model}\label{Proca}
\subsection{Green function expansions}{\label{Greenfunctionexpansions}

The Lagrangian density for a vector field of mass $m$ will be taken as
\begin{equation}\label{Lagrangian}
{\mathcal L}=\frac{1}{4}F_{\mu\nu}F^{\mu\nu}+\frac{1}{2}m^2A^\mu A_\mu+\frac{1}{2}X^{\mu\nu}A_\mu A_\nu\;.
\end{equation}
Here $F_{\mu\nu}=\nabla_\mu A_\nu-\nabla_\nu A_\mu$ as usual. In this section we will adopt a Riemannian metric and our curvature conventions will be those of \cite{MTW}. 

Normally the last term involving $X^{\mu\nu}$ is not present. We will refer to a non-zero value of $X^{\mu\nu}$ as a non-minimally coupled vector field. The case of $X^{\mu\nu}=0$ will be called the minimal vector field and corresponds to the original Proca model. We will examine the reason for this terminology, and the connection with its previous usage as in \cite{BarvinskyVilkovisky} for example, later. The presence of $X^{\mu\nu}$ means that the theory is not gauge invariant even if the mass term $m^2=0$ is chosen. So this allows us to consider a photon that has a nonminimal coupling to the curvature for example. By analogy with the usual scalar field $R\phi^2$ coupling we can have the special case
\begin{equation}
X_{\mu\nu}=\xi_1\,R_{\mu \nu}+\xi_2\, R g_{\mu\nu}\;,\label{nonmin}
\end{equation}
where $\xi_1$ and $\xi_2$ are dimensionless coupling constants although we will keep $X_{\mu\nu}$ general at this stage, assuming only without loss of generality that it is symmetric.

Because there is no gauge invariance we would expect that we can express the one-loop effective action as
\begin{equation}
\Gamma^{(1)}=\frac{1}{2}\ln\det\Delta^{\mu}{}_{\nu}\,\label{oneloop}
\end{equation}
where
\begin{equation}
\Delta^{\mu}{}_{\nu}=-\delta^{\mu}_{\nu}\Box+\nabla^\mu\nabla_\nu+R^{\mu}{}_{\nu}+m^2\delta^{\mu}_{\nu}+X^{\mu}{}_{\nu}\;.\label{delta1}
\end{equation}
This can be read off directly from \eqref{Lagrangian}. We will examine this critically later where we will show, perhaps surprisingly, that this is actually only true in the minimal case. Nevertheless the analysis of the heat kernel coefficients for the operator $\Delta^{\mu}{}_{\nu}$ that we will give holds regardless of whether or not \eqref{oneloop} with \eqref{delta1} is really the effective action or not; $\ln\det\Delta^{\mu}{}_{\nu}$ has a well defined meaning regardless of its physical interpretation.

In the case $X_{\mu\nu}=0$ Barvinsky and Vilkovisky~\cite{BarvinskyVilkovisky} show that
\begin{equation}
\Delta^{\mu}{}_{\lambda}\big(\delta^{\lambda}_{\nu}-\frac{1}{m^2}\nabla^\lambda\nabla_\nu\big)=\delta^{\mu}_{\nu}(-\Box+m^2)+R^{\mu}{}_{\nu}\;.\label{BVidentity}
\end{equation}
This allows the operator $\Delta^{\mu}{}_{\nu}$ to be related to the simpler operator on the right hand side whose leading order derivative is $\Box$. This clever identity therefore relates the nonminimal operator $\Delta^{\mu}{}_{\nu}$ that appears in \eqref{delta1} to the minimal operator on the right hand side of \eqref{BVidentity} resulting in considerable simplification in the calculation of one loop divergences and heat kernel coefficients~\cite{BarvinskyVilkovisky}. If the operator has a non-zero value for $X_{\mu\nu}$ then this identity no longer holds and the calculation is more involved. It is possible to obtain a modified version of \eqref{BVidentity} when $X^{\mu\nu}$ is nonzero, but this turns out not to have the same utility that \eqref{BVidentity} has for $X^{\mu\nu}=0$.

Since the Green's function is the formal inverse of the operator, we have
\begin{equation}
\frac{\partial}{\partial m^2}\Gamma^{(1)}=\frac{1}{2}\int dv_x\,G^{\mu}{}_{\mu}(x,x)\,\label{gammmaderiv}
\end{equation}
where 
\begin{equation}
\Delta^{\mu}{}_{\lambda}G^{\lambda}{}_{\nu}(x,x')=\delta^{\mu}_{\nu}\delta(x,x')\;.\label{Greenfcn}
\end{equation}
We can now use the local momentum space expansion to calculate $G^{\mu}{}_{\nu}(x,x')$. Most of the work was done in \cite{toms2014local} and we refer to this reference for a detailed description of the method. To facilitate the comparison with results from \citep{toms2014local} we will define
\begin{equation}
\Delta^{\mu}{}_{\nu}=\delta^{\mu}_{\nu}(-\Box+m^2)+\nabla^\mu\nabla_\nu+Q^{\mu}{}_{\nu}\;,\label{delta}
\end{equation}
where 
\begin{equation}
Q^{\mu}{}_{\nu}=R^{\mu}{}_{\nu}+X^{\mu}{}_{\nu}\;.\label{Q}
\end{equation}
We will develop expressions valid for any $Q^{\mu}{}_{\nu}$. The local momentum space expressions can now be recovered directly from those of \citep{toms2014local} by letting $q\rightarrow1$ and $s\rightarrow (-m^2)$ in the expression referred to as the auxiliary Green function.

The results for the local momentum expansion turn out to be
\begin{equation}
(G_0)_{\mu \nu}=\left(\delta_{\mu\nu}+\frac{p_{\mu}p_{\nu}}{m^2}\right)\,S\;,\label{G0}
\end{equation}
where we have defined
\begin{equation}
S=(p^2+m^2)^{-1}\;.\label{S}
\end{equation}
It is not necessary to refer the components to a local orthonormal frame as in \cite{toms2014local} because we will only consider the coincidence limit in this paper where this distinction disappears. The result of \eqref{G0} is recognized as the usual flat spacetime expression for the Green function in momentum space. (The expression referred to as $T$ in \citep{toms2014local} becomes just $T=1/m^2$ here with the above-mentioned replacements. This is why it is advantageous to include the $m^2$ term in $G_0$ rather than as part of $X^{\mu}{}_{\nu}$ or equivalently $Q^{\mu}{}_{\nu}$.)

The next term in the local momentum space expansion is
\begin{eqnarray}
{(G_2)}_{\mu \nu} &=&\frac{1}{3}\, R \left\lbrack {\delta}_{\mu\nu}+\frac{ {p}_{\mu} {p}_{\nu}}{m^2}  \right\rbrack\,S{}^{2} - \frac{1}{3}\, \frac{{p}_{\alpha} {p}_{\beta}}{m^2} \,{R}_{\mu}{}^{ \alpha}{}_{\nu}{}^{ \beta}\, S{}^{2} \nonumber\\ 
&&-  \left\lbrack {Q}_{\mu\nu} +\frac{1}{m^2}\,Q_{\mu\lambda}\,{p}^{\lambda} {p}_{\nu} +\frac{1}{m^2}\,Q_{\nu\lambda}\,{p}^{\lambda} {p}_{\mu}   + \frac{1}{m^4}\,Q_{\alpha\beta}\,{p}^{\alpha} {p}^{\beta} {p}_{\mu} {p}_{\nu}    \right\rbrack\,S^2\nonumber\\
&&   - \frac{2}{3}\,{R}_{\alpha\beta}\, {p}^{\alpha} {p}^{\beta} S{}^{3} {\delta}_{\mu\nu} +\frac{1}{m^2}\,R_{\mu\lambda}\,{p}^{\lambda} {p}_{\nu}\,S^2 +\frac{1}{m^2}\,R_{\nu\lambda}\,{p}^{\lambda} {p}_{\mu}\,S^2\nonumber\\
&&\qquad + \frac{1}{6m^2}\, R_{ \mu \nu}\, S  +  \frac{1}{m^4}\,R_{\alpha\beta}\,{p}^{\alpha} {p}^{\beta} {p}_{\mu} {p}_{\nu} \big( - \frac{2m^2}{3}\, S  + 1 \big)\,S^2 \;.\label{G2}
\end{eqnarray}
The full expression here is needed in order to compute $(G_4)_{\mu \nu}$. However for the effective action we just need the trace of \eqref{G2}. This is easily calculated to be (with $N$ the spacetime dimension)
\begin{eqnarray}
{\rm tr}(G_2) &=& - {Q}\, S{}^{2} - {Q}^{\alpha \beta}\, \frac{{p}_{\alpha} {p}_{\beta}}{m^2} \left\lbrack S^2 + \frac{1}{m^2}\,S\right\rbrack\nonumber\\
&& + R \left\lbrack\frac{(N-1)}{3}\,S{}^{2} + \frac{1}{2m^2}\, S  \right\rbrack + {R}^{\alpha \beta}\, {p}_{\alpha} {p}_{\beta} \left\lbrack  \frac{2(1-N)}{3}\,  S{}^{3}  + \frac{1}{m^4}\,S \right\rbrack\;,\label{trG2}
\end{eqnarray}
where we have abbreviated
\begin{equation}
Q=Q^{\alpha}{}_{\alpha}\;.\label{trQ}
\end{equation}

The term $(G_3)_{\mu\nu}$ is easy to determine, but we will not need it here so shall not quote it for brevity. Because it is odd in the momentum it makes to contribution to the effective action.

Instead of quoting the full expression for $(G_4)_{\mu\nu}$ that we will not need, we will just give the result for its trace that is needed for the one-loop divergent part of the effective action and the trace of the heat kernel coefficients. We will give the result that is found for $(G_4)_{\mu\nu}(x',x')$ after performing the momentum integrations below.
\begin{eqnarray}
{\rm tr}(G_4)&=& - \frac{2}{3}\, Q R\, S{}^{3} + R{}^{2} \left\lbrack  \frac{(N-1)}{9}\, S{}^{3}+ \frac{1}{6m^2}\,  S{}^{2}\right\rbrack\nonumber\\
&& -{Q}_{\alpha \beta} R \left\lbrack \frac{2}{3}\,  S{}^{3} + \frac{5}{6m^2}\, S{}^{2} \right\rbrack \frac{{p}^{\alpha} {p}^{\beta}}{m^2} + {Q}_{\alpha \beta} {Q}^{\alpha \beta}\, S{}^{3}\nonumber\\
&& + {Q}^{\alpha}{}_{ \beta} {Q}_{\alpha \mu} \left\lbrack2\,  S{}^{3} + \frac{1}{m^2}\,S{}^{2} \right\rbrack\frac{ {p}^{\beta} {p}^{\mu}}{m^2} + {Q}_{\alpha \beta} {Q}_{\mu \nu} \left\lbrack S{}^{3} + \frac{1}{m^2}S{}^{2} \right\rbrack\frac{ {p}_{\alpha} {p}_{\beta} {p}_{\mu} {p}_{\nu}}{m^4}\nonumber\\
&& + 2\,{R}_{\alpha \beta}\,Q\, S{}^{4} {p}^{\alpha} {p}^{\beta} 
+{R}_{\alpha \beta} R \left\lbrack\frac{2(1-N)}{3}\, S{}^{4}  - \frac{1}{3m^2}\,  S{}^{3} + \frac{5}{6m^4}\, S{}^{2} \right\rbrack {p}^{\alpha} {p}^{\beta}\nonumber\\
&& +{R}_{\alpha \beta} {Q}^{\alpha \beta} \left\lbrack \frac{1}{3m^4}\, S  - \frac{2}{3m^2}\,  S{}^{2}\right\rbrack - {Q}^{\alpha}{}_{ \beta} {R}_{\alpha \mu} \left\lbrack  S{}^{3} + \frac{7}{2m^2}\, S{}^{2} \right\rbrack \frac{{p}^{\beta} {p}^{\mu}}{m^2}\nonumber\\
&& + {Q}_{\alpha \beta} {R}_{\mu \nu} \left\lbrack 2\,  S{}^{4} -\frac{ 2}{m^4}\, S{}^{2}  + \frac{2}{3m^2}\, S{}^{3} \right\rbrack\frac{ {p}^{\alpha} {p}^{\beta} {p}^{\mu} {p}^{\nu}}{m^2}\nonumber\\
&& + {R}_{\alpha \beta} {R}^{\alpha \beta} \left\lbrack  \frac{2(N-1)}{45}\, S{}^{3}+ \frac{3}{4m^2}\, S{}^{2}  - \frac{1}{3m^4}\, S  \right\rbrack\nonumber\\
&& + {R}^{\alpha}{}_{ \beta} {R}_{\alpha \mu} \left\lbrack \frac{(13-8N)}{5}\, S{}^{4} - \frac{3}{m^2}\,  S{}^{3} + \frac{3}{m^4}\, S{}^{2} \right\rbrack {p}^{\beta} {p}^{\mu}\nonumber\\
&& + {R}_{\alpha \beta} {R}_{\mu \nu} \left\lbrack  \frac{4(N-1)}{3}\, S{}^{5}  - \frac{5}{3m^4}\, S{}^{3}  + \frac{1}{m^6}\,S{}^{2}\right\rbrack {p}^{\alpha} {p}^{\beta} {p}^{\mu} {p}^{\nu}\nonumber\\
&& -\frac{1}{3}\, {R}_{\alpha \mu \beta \nu} {Q}^{\alpha \beta} \left\lbrack S{}^{3}+  \frac{1}{2m^2}\, S{}^{2} \right\rbrack\frac{ {p}^{\mu} {p}^{\nu}}{m^2}\nonumber\\
&& + {R}_{\alpha \mu \beta \nu}{R}^{\alpha \beta} \left\lbrack \frac{(4N-9)}{5}\, S{}^{4} + \frac{1}{m^2}\,S{}^{3} - \frac{1}{3m^4}\, S{}^{2} \right\rbrack {p}^{\mu} {p}^{\nu}\nonumber\\
&& + {R}^{\alpha \beta \lambda}{}_{ \mu } {R}_{\alpha \beta \lambda \nu} \left\lbrack  \frac{2(19-4N)}{15}\, S{}^{4} - \frac{1}{m^2}\, S{}^{3}\right\rbrack{p}^{\mu} {p}^{\nu}\nonumber\\
&& + {R}^{\alpha \beta \mu \nu} {R}_{\alpha \beta \mu \nu} \left\lbrack \frac{(2N-17)}{30}\, S{}^{3}  + \frac{1}{8m^2}\,  S{}^{2}\right\rbrack\nonumber\\
&& + {R}^{\alpha}{}_{ \beta}{}^{ \mu}{}_{ \nu} {R}_{\alpha \lambda \mu \sigma} \left\lbrack \frac{16(N-1)}{15}\, S{}^{5}+\frac{2 }{m^2}\, S{}^{4}\right\rbrack {p}^{\beta} {p}^{\lambda} {p}^{\nu} {p}^{\sigma}\;.\label{trG4}
\end{eqnarray}

To do the integration over momentum we use the standard result
\begin{equation}
\int\frac{d^Np}{(2\pi)^N}\,(p^2+m^2)^{-n}\,p_{\mu_1}\cdots p_{\mu_{2k}}=I_{k}(n)\delta_{\mu_1\cdots\mu_{2k}}\;,\label{momint}
\end{equation}
where
\begin{equation}
I_{k}(n)=\frac{1}{2^k}(4\pi)^{-N/2}(m^2)^{N/2-n+k}\,\frac{\Gamma(n-k-N/2)}{\Gamma(n)}\;,\label{Ik}
\end{equation}
and $\delta_{\mu_1\cdots\mu_{2k}}$ is the sum over all symmetrized products of Kronecker deltas with all possible index pairings. Here $k=0,1,2,\ldots$. 

For the first term $G_0$ the result is found from \eqref{G0} to be
\begin{equation}
{\rm tr}G_0(x',x')=(N-1)I_0(1).\label{G0e}
\end{equation}
In obtaining this expression it is necessary to discard the divergence that is quartic in the case of $N=4$. This can be justified using dimensional regularization by the usual rule where such a divergence is proportional to $\delta(0)$ and this is regularized to zero. As we will see in Sec.~\ref{sec_Procaquantized} when we analyze the quantization of the model more closely this regularization is not needed. 

For $G_2(x',x')$ we can use \eqref{G2} to find 
\begin{equation}
{\rm tr}G_2(x',x')=\frac{1}{4}(4\pi)^{-N/2}\Gamma\Big(-\frac{N}{2}\Big)(m^2)^{N/2-2}\big\lbrack \frac{1}{6}\,(N^2-4)(N-3)\,R-(N-1)(N-2)\,Q \big\rbrack.\label{G2e}
\end{equation}

The result for $(G_{4})_{\mu \nu}(x',x')$ is
\begin{eqnarray}
(4\pi)^{N/2}\,(G_{4})_{\mu \nu}(x',x')&=& 
t_{41}\,Q^{\alpha \beta}{}_{;\alpha\beta} \delta_{\mu \nu} 
+t_{42}\, Q_{\nu}{}^{ \alpha}{}_{;\alpha\mu}
+t_{42}\, Q_{\mu}{}^{ \alpha}{}_{;\alpha\nu}  
+t_{42}\, Q_{\nu}{}^{ \alpha}{}_{;\mu\alpha} 
\nonumber\\
&&+t_{42} \,Q_{\mu}{}^{ \alpha}{}_{;\nu\alpha}
+ t_{48}\,\Box\,Q_{\mu \nu}
+ t_{49}\,\Box\,R_{\mu \nu} 
+ t_{410}\,\Box\,Q\, g_{\mu \nu}  \nonumber\\
&&+2\,t_{410}\,Q_{;\mu\nu} 
+t_{412}\,\Box\,R\, g_{\mu \nu} 
+t_{413}\, R_{;\mu\nu} 
+t_{414}\, R^{2} g_{\mu \nu} 
\nonumber\\
&&+t_{415}\, Q^{2} g_{\mu \nu}+t_{416}\, Q Q_{\mu \nu} 
+t_{417}\, Q R_{\mu \nu} 
+t_{418}\, Q^{\alpha \beta} Q_{\alpha \beta} g_{\mu \nu} 
 \nonumber\\
&&+t_{419}\, Q_{\mu}{}^{ \alpha} Q_{\nu \alpha}
+t_{420}\, Q^{\alpha \beta} R_{\mu \alpha \nu \beta}+t_{421}\, R Q g_{\mu \nu} 
+t_{422}\, R Q_{\mu \nu}  
\nonumber\\
&&+t_{423}\, R R_{\mu \nu}+t_{424}\, R^{\alpha \beta} Q_{\alpha \beta} g_{\mu \nu} 
+t_{425}\, R_{\mu}{}^{ \alpha} Q_{\nu \alpha} +t_{425}\, R_{\nu}{}^{ \alpha} Q_{\mu \alpha} 
\nonumber\\
&&+t_{427}\, R^{\alpha \beta} R_{\alpha \beta} g_{\mu \nu}+t_{428}\, R_{\mu}{}^{ \alpha} R_{\nu \alpha} 
+t_{429}\, R^{\alpha \beta} R_{\mu \alpha \nu \beta}\nonumber\\
&&+t_{430}\, R^{\alpha \beta \lambda \sigma} R_{\alpha \beta \lambda \sigma} g_{\mu \nu}
+t_{431}\, R_{\mu}{}^{ \alpha \beta \lambda} R_{\nu \alpha \beta \lambda} \;,\label{G4munu}
\end{eqnarray}
where
\begin{eqnarray}
t_{41}&=&-\frac{1}{12}\, m^{N-6} \Gamma \left(1-\frac{N}{2}\right) \nonumber\\
t_{42}&=&\frac{1}{24}\, (5-2 N)\, m^{N-6}\, \Gamma \left(1-\frac{N}{2}\right)\nonumber\\
t_{48}&=&\frac{1}{48}\, N (N^2-8N+14)\, m^{N-6}\, \Gamma \left(-\frac{N}{2}\right)\nonumber\\
t_{49}&=&\frac{1}{120}\, (28-9 N)\, m^{N-6}\, \Gamma \left(1-\frac{N}{2}\right)\nonumber\\
t_{410}&=&-\frac{1}{24}\, m^{N-6}\, \Gamma \left(1-\frac{N}{2}\right)\nonumber\\
t_{412}&=&\frac{1}{120}\, (N^2-7N+20)\, m^{N-6}\, \Gamma \left(1-\frac{N}{2}\right)\nonumber\\
t_{413}&=&-\frac{3}{10}\,  m^{N-6}\, \Gamma \left(2-\frac{N}{2}\right)\nonumber\\
t_{414}&=&-\frac{1}{576}\, (N^3-7N^2+22N-36)\, m^{N-6}\, \Gamma \left(-\frac{N}{2}\right)\nonumber\\
t_{415}&=&\frac{1}{16}\, m^{N-6}\, \Gamma \left(-\frac{N}{2}\right)\nonumber\\
t_{416}&=&-\,\frac{m^{N-6}\, \Gamma \left(2-\frac{N}{2}\right)}{2 N}\nonumber\\
t_{417}&=&\frac{1}{12}\, (N-3)\, m^{N-6}\, \Gamma \left(-\frac{N}{2}\right)\nonumber\\
t_{418}&=&-\,\frac{m^{N-6}\, \Gamma \left(2-\frac{N}{2}\right)}{4 N}\nonumber\\
t_{419}&=&-\,\frac{1}{16}\, (N^3-9N^2+20N-8)\, m^{N-6}\, \Gamma \left(-\frac{N}{2}\right)\nonumber\\
t_{420}&=&-\,\frac{1}{12}\, m^{N-6}\, \Gamma \left(1-\frac{N}{2}\right)\nonumber\\
t_{421}&=&\frac{1}{48}\, (N-6)\, m^{N-6}\, \Gamma \left(-\frac{N}{2}\right)\nonumber\\
t_{422}&=&\frac{1}{48}\, (N^3-8N^2+20N-12)\, m^{N-6}\, \Gamma \left(-\frac{N}{2}\right)\nonumber\\
t_{423}&=&\frac{(5 N-18)\, m^{N-6}\, \Gamma \left(2-\frac{N}{2}\right)}{36 N}\nonumber\\
t_{424}&=&\frac{1}{12}\, (N-3)\, m^{N-6}\, \Gamma \left(-\frac{N}{2}\right)\nonumber\\
t_{425}&=&-\,\frac{1}{48}\, (N-1) (7 N-24)\, m^{N-6}\, \Gamma \left(-\frac{N}{2}\right)\nonumber\\
t_{427}&=&\frac{(N^3-7N^2-20N+180)\, m^{N-6}\, \Gamma \left(-\frac{N}{2}\right)}{1440}\nonumber\\
t_{428}&=&\frac{1}{720} (19 N^2-248N+360)\, m^{N-6}\, \Gamma \left(-\frac{N}{2}\right)\nonumber\\
t_{429}&=&\frac{1}{180}\, (88-29 N)\, m^{N-6}\, \Gamma \left(1-\frac{N}{2}\right)\nonumber\\
t_{430}&=&-\,\frac{1}{360}\, (N-5)\, m^{N-6}\, \Gamma \left(2-\frac{N}{2}\right)\nonumber\\
t_{431}&=&\frac{1}{360}\, (15 N-64)\, m^{N-6}\, \Gamma \left(2-\frac{N}{2}\right)\;,
\end{eqnarray}

We can work out the pole part of this expression at $N=6$. This results in
\begin{eqnarray}
(G_{4})_{\mu \nu}(x',x')&=& \frac{1}{(4\pi)^3(N-6)}\Big\lbrack
\frac{1}{12}\,Q^{\alpha \beta}{}_{;\alpha\beta} \delta_{\mu \nu} 
+\frac{7}{24}\, Q_{\nu}{}^{ \alpha}{}_{;\alpha\mu}
+\frac{7}{24}\, Q_{\mu}{}^{ \alpha}{}_{;\alpha\nu}  
+\frac{7}{24}\, Q_{\nu}{}^{ \alpha}{}_{;\mu\alpha} 
\nonumber\\
&&+\frac{7}{24} \,Q_{\mu}{}^{ \alpha}{}_{;\nu\alpha}
+ \frac{1}{12}\,\Box\,Q_{\mu \nu}
+ \frac{13}{60}\,\Box\,R_{\mu \nu} 
+ \frac{1}{24}\,\Box\,Q\, g_{\mu \nu}  \nonumber\\
&&+\frac{1}{12}\,Q_{;\mu\nu} 
-\frac{7}{60}\,\Box\,R\, g_{\mu \nu} 
-\frac{3}{5}\, R_{;\mu\nu} 
-\frac{5}{144}\, R^{2} g_{\mu \nu} 
\nonumber\\
&&+\frac{1}{48}\, Q^{2} g_{\mu \nu}
-\frac{1}{6}\, Q Q_{\mu \nu} 
+\frac{1}{12}\, Q R_{\mu \nu} 
-\frac{1}{12}\, Q^{\alpha \beta} Q_{\alpha \beta} g_{\mu \nu} 
 \nonumber\\
&&-\frac{1}{12}\, Q_{\mu}{}^{ \alpha} Q_{\nu \alpha}
+\frac{1}{12}\, Q^{\alpha \beta} R_{\mu \alpha \nu \beta}
+\frac{1}{4}\, R Q_{\mu \nu}  
\nonumber\\
&&+\frac{1}{9}\, R R_{\mu \nu}
+\frac{1}{12}\, R^{\alpha \beta} Q_{\alpha \beta} g_{\mu \nu} 
-\frac{5}{8}\, R_{\mu}{}^{ \alpha} Q_{\nu \alpha} 
-\frac{5}{8}\, R_{\nu}{}^{ \alpha} Q_{\mu \alpha} 
\nonumber\\
&&+\frac{1}{180}\, R^{\alpha \beta} R_{\alpha \beta} g_{\mu \nu}
-\frac{37}{180}\, R_{\mu}{}^{ \alpha} R_{\nu \alpha} 
+\frac{43}{90}\, R^{\alpha \beta} R_{\mu \alpha \nu \beta}\nonumber\\
&&-\frac{1}{180}\, R^{\alpha \beta \lambda \sigma} R_{\alpha \beta \lambda \sigma} g_{\mu \nu}
+\frac{13}{90}\, R_{\mu}{}^{ \alpha \beta \lambda} R_{\nu \alpha \beta \lambda}\Big\rbrack \;,\label{nequals6}
\end{eqnarray}
If the heat kernel coefficients are not dependent on the spacetime dimension then this should be simply related~\cite{TomsPRDscalar} to $E_2$. In a similar way the pole for $N=4$ should be related to the $E_1$ coefficient. (This will be shown in the next section.)  We can check this result against the other independent expression for the divergent part of the effective action at least with the total derivatives excluded. If we take the trace of the expression with $N=4$ we do not end up with a result that agrees with the pole part of the effective action below. We conclude that the heat kernel coefficients must depend on the spacetime dimension. These will be determined below.

Taking the trace of the $(G_{4})_{\mu \nu}(x',x')$ expression, and dropping the terms that are total derivatives, gives
\begin{eqnarray}
(4\pi)^{N/2}\,(G_{4})^{\mu}{}_{ \mu}(x',x')&=& 
(N t_{414}+t_{423})\, R^{2} 
+ (N t_{415}+t_{416})\, Q^{2} 
+(t_{417}+N t_{421}+t_{422})\, Q R \nonumber\\
&&
+(N t_{418}+t_{419})\, Q^{\alpha \beta} Q_{\alpha \beta} 
+(t_{420}+N t_{424}+2\,t_{425})\, Q^{\alpha \beta} R_{\alpha \beta}\nonumber\\
&&
+(N t_{427}+t_{428}+t_{429})\, R^{\alpha \beta} R_{\alpha \beta} 
+(t_{430}+t_{431})\, R^{\alpha \beta \lambda \sigma} R_{\alpha \beta \lambda \sigma}, \label{G4e}
\end{eqnarray}

We can now make use \eqref{gammmaderiv} along with the results found in \eqref{G0e},\eqref{G2e} and \eqref{G4e} to find the pole part of $\Gamma^{(1)}$ for $N=4$. It follows that 
\begin{eqnarray}
{\rm PP}(\Gamma^{(1)})&=&\frac{1}{16\pi^2(N-4)}\int dv_x \Big\lbrack \frac{1}{48}\,R^2 +\frac{1}{16}\,Q^2 + \frac{1}{8}\,Q_{\alpha\beta}Q^{\alpha\beta}-\frac{5}{24}\,RQ\nonumber\\
&&\qquad +\frac{7}{12}\,R^{\mu \nu}Q_{\mu\nu} - \frac{9}{40}\,R^{\mu\nu}R_{\mu\nu} -\frac{1}{15} \, R^{\alpha \beta \lambda \sigma} R_{\alpha \beta \lambda \sigma}\nonumber\\
&&\qquad+\frac{3\,m^2}{4}\,Q-\frac{m^2}{4}\,R+\frac{3}{2}\,m^4
\big\rbrack\;.\label{onelooppole1}
\end{eqnarray}
Here ${\rm PP}$ denotes the pole part of any expression. If we specialize to the case $X_{\mu\nu}=0$, so that $Q_{\mu\nu}=R_{\mu\nu}$ then we find
\begin{eqnarray}
{\rm PP}(\Gamma^{(1)})&=&\frac{1}{16\pi^2(N-4)}\int dv_x \Big\lbrack -\frac{1}{8}\,R^2  + \frac{29}{60}\,R^{\mu\nu}R_{\mu\nu} -\frac{1}{15} \, R^{\alpha \beta \lambda \sigma} R_{\alpha \beta \lambda \sigma}\nonumber\\
&&\qquad+\frac{m^2}{2}\,R+\frac{3}{2}\,m^4
\Big\rbrack\;.\label{onelooppole}
\end{eqnarray}
This is in complete agreement with equation (5.48) of \citep{BarvinskyVilkovisky}. Note that we have established this with a direct calculation based on the operator $\Delta^{\mu}{}_{\nu}$ in its original form without any use of the identity \eqref{BVidentity} that was established in \citep{BarvinskyVilkovisky}.

\subsection{Heat kernel and Green function relations}\label{sec-heat}

The relationship between the Green function and the heat kernel is (based on the simple observation that the Green function is the formal inverse of the differential operator that defines it):
\begin{equation}
G(x,x)=\int\limits_{0}^{\infty}d\tau\,e^{-m^2\,\tau}\,\tilde{K}(x,x;\tau)\;,\label{HK1}
\end{equation}
where the tilde denotes the heat kernel for the operator $\Delta-m^2\,I$ ({\em i.e.}\/ with the $m^2$ term removed). Based on experience with other operators (see for example \cite{DeWittdynamical}) we might now assume that the Schwinger-DeWitt asymptotic expansion for the heat kernel provides us with a local curvature expansion for the Green function. By adopting dimensional regularization, and assuming that the local curvature expansion comes from the asymptotic expansion
\begin{equation}
\tilde{K}(x,x;\tau)\simeq(4\pi\tau)^{-N/2}\,\sum\limits_{k=0}^{\infty}\tau^k\,E_k(x)\;,\label{HK2}
\end{equation}
it is easily shown that
\begin{equation}
G(x,x)\simeq(4\pi)^{-N/2}\,\sum\limits_{k=0}^{\infty}(m^2)^{N/2-k-1}\Gamma(k+1-N/2)\,\,E_k(x)\;. \label{HK3}
\end{equation}
In our conventions $G_{2k}(x,x)$ denotes that part of the Green function that contains terms that are of the order $k$ in the curvature. Since $E_k$ also contains the terms of order $k$ in the curvature we have the correspondence
\begin{equation}
G_{2k}(x,x)=(4\pi)^{-N/2}\,(m^2)^{N/2-k-1}\Gamma(k+1-N/2)\,\,E_k(x)\;.\label{HK4}
\end{equation}
Since our local momentum space expansion is used to evaluate $G_{2k}(x,x)$, at least for $k=0,1,2$ we can use the results to obtain the first few heat kernel coefficients. We have checked that this agrees with the more elaborate procedure of using the auxiliary Green function method described in our earlier work \cite{MossToms,toms2014local} in the case of minimal operators for scalars and the nonminimal operator for vectors. Although it is possible to obtain the untraced heat kernel coefficients we will concentrate on just the traced expressions as these are the ones relevant for the effective action.

If we take $k=0$ in \eqref{HK4} and use \eqref{G0e} with \eqref{Ik} it can be seen that
\begin{equation}
{\rm tr}\,E_0=N-1.\label{trE0}
\end{equation}
In obtaining this the $\delta(0)=0$ result has been used.

If we take $k=1$ in \eqref{HK4} we may use \eqref{G2e} with \eqref{Ik} to find
\begin{equation}
{\rm tr}\,E_1=\frac{(N+2)(N-3)}{6N}\,R +\left(1-\frac{1}{N}\right)\,Q.\label{trE1}
\end{equation}

In principle we should be able to use the expression for \eqref{G4munu} and the associated coefficients listed below to read off $(E_2)_{\mu\nu}(x)$ or its trace. However if this is done the result is found to diverge as we try to let $N\rightarrow4$. We are therefore not able to deduce the untraced or traced heat kernel coefficients in this way. It is obvious to try to use the auxiliary Green function approach used in \cite{MossToms,toms2014local}; however this too runs into difficult. In this approach the operator \eqref{delta1} results in expressions that have divergent momentum integrations. The root cause of these problems is the second term in the flat spacetime momentum space Green function \eqref{BVidentity}. The fact that there will be problems should be evident from the nonminimal vector results that we obtained in \cite{MossToms,toms2014local} since \eqref{delta1} is related to the limit $q\rightarrow1$ of these results and this limit does not exist in general. This problem persists even if we evaluate the trace of $E_2$ and keep the total derivative terms. Fortunately in our application we do not need total derivative terms. If we concentrate on just the trace and discard total derivatives we find
\begin{eqnarray}
{\rm tr}\lbrack E_2(x)\rbrack&=&t_{21}\,R^2+t_{22}\,Q^2+t_{23}\,RQ+t_{24}\,Q^{\mu\nu}Q_{\mu\nu} +t_{25}\,R^{\mu\nu}Q_{\mu\nu}\nonumber\\
&&+t_{26}\,R^{\mu\nu}R_{\mu\nu}+t_{27}\,R^{\mu\nu\lambda\sigma}R_{\mu\nu\lambda\sigma}\;,\label{HK5}
\end{eqnarray}
where
\begin{eqnarray}
t_{21}&=&\frac{(36-10\,N-3\,N^2+N^3)}{72\,N(N-2)}\;,\label{HK6a}\\
t_{22}&=&\frac{1}{2\,N(N-2)}\;,\label{HK6b}\\
t_{23}&=&-\,\frac{(N^2-3\,N+6)}{6\,N(N-2)}\;,\label{HK6c}\\
t_{24}&=&\frac{(N^2-4\,N+2)}{2\,N(N-2)}\;,\label{HK6d}\\
t_{25}&=&\frac{(5\,N-6)}{3\,N(N-2)}\;,\label{HK6e}\\
t_{26}&=&-\,\frac{(N^3-3\,N^2+122\,N-180)}{180\,N(N-2)}\;,\label{HK6f}\\
t_{27}&=&\frac{(N-16)}{180}\;.\label{HK6g}
\end{eqnarray}
It is easy to check that if we let $N\rightarrow4$ the results are in complete agreement with the result of \eqref{onelooppole1}. If we let $N\rightarrow6$ then we find results that agree with the pole determination of \eqref{nequals6}. We have therefore found the coefficients for any spacetime dimension other than 2. By making use of \eqref{Q} and using the results in \eqref{HK6a}--\eqref{HK6g} we find
\begin{eqnarray}
{\rm tr}\lbrack E_2(x)\rbrack&=&\frac{(N-13)}{72}\,R^2+t_{22}\,S^2-\frac{1}{6}\,\left(\frac{N-3}{N-2}\right)\,RX+t_{24}\,X^{\mu\nu}X_{\mu\nu} \label{HK7}\\
&&+\frac{1}{3}\,\left(\frac{3\,N-7}{N-2}\right)\,R^{\mu\nu}X_{\mu\nu}-\left(\frac{N-91}{180}\right)\,R^{\mu\nu}R_{\mu\nu}+\left(\frac{N-16}{180}\right)\,R^{\mu\nu\lambda\sigma}R_{\mu\nu\lambda\sigma}.\nonumber
\end{eqnarray}

We have therefore found the first few heat kernel coefficients in a general spacetime dimension but without calculating the terms that are total derivatives. We will return to the total derivative terms in Sec.~\ref{sec-ultra}.

\section{Quantization of the Proca field}\label{sec_Procaquantized}

If the one-loop effective action for the Proca field was given by \eqref{oneloop} and \eqref{delta1} then the paper would have stopped with the last section. However we will now show that this is only true for the minimal Proca model where $X^{\mu\nu}=0$. We will show that it is not true for the generalized Proca model. In fact we will show that in general if we allow a nonzero value for $X^{\mu\nu}$ we end up with a theory that is not renormalizable. At first sight this might seem a bit surprising as adding on the last term to \eqref{Lagrangian} looks like a very modest modification to the original Proca theory. It turns out however that the situation is much more involved than it appears. In the next subsection, Sec.~\ref{sec-FJ}, we will discuss the correct quantization procedure for the generalized Proca model. In the following subsection, Sec.~\ref{sec-ultra}, we will show how this quantization procedure recovers the results of Sec~\ref{Proca} when we take $X^{\mu\nu}=0$, so that the results found above, and earlier by Barvinsky and Vilkovisky~\citep{BarvinskyVilkovisky} are correct in this special case. In the final subsection, Sec.~\ref{sec-flat}, we will specialize to flat spacetime and consider two special cases of $X^{\mu\nu}$ to demonstrate that the modified Proca theory is not renormalizable in general.

\subsection{Faddeev-Jackiw quantization of the generalized Proca model}\label{sec-FJ}

It turns out, as mentioned, that the one-loop effective action is not really given by \eqref{oneloop} in general. This would certainly be true if all of the components of the vector field were independent degrees of freedom. Of course for the Proca field this is not the case as the system is constrained. In the Lagrangian of \eqref{Lagrangian} there are only time derivatives of the spatial components of the vector field; the time component $A_0$ appears as an auxiliary field without any time derivatives. This is of course well-known in flat spacetime~\cite{Wentzel} and the situation in curved spacetime is no different. There are a variety of different ways to proceed, the most common being that due to Dirac~\cite{DiracLecturesQM}. For a variety of different approaches see for example~\cite{Sundermeyer,GitmanTyutin,HenneauxTeitelboim,RotheandRothe} and references therein. In this section we will use a very elegant method introduced by Faddeev and Jackiw~\cite{FaddeevJackiw} along with the path integral method to consider the quantization of the Proca field. A good review of the method and a discussion of the Proca field in flat spacetime is given in \cite{Jackiwnotears}.

Because we will be separating off the time derivatives here it is advantageous to adopt a metric signature of $(-,+,\cdots,+)$ for this section in place of the Riemannian signature used up until now. We will assume the spacetime dimension to be $N$ as before, but now for simplicity will restrict the spacetime to be ultrastatic (see~\cite{fulling1989aspects} for example) meaning that we take the line element to be
\begin{equation}
ds^2=-dt^2+g_{ij}({\mathbf x})\,dx^i\,dx^j\;.\label{ultrastatic}
\end{equation}
The spatial metric $g_{ij}$ is assumed to be Riemannian with indices $i,j,\ldots$ running over the $(N-1)$ spatial dimensions. It is possible to be more general than this, but to do so adds unnecessary complexity to the points we are trying to make. (See \citep{prescod2015extension} for example.)

With the Lorentzian signature that we have adopted in this section the Lagrangian density becomes the negative of the Riemannian one in \eqref{Lagrangian} and is
\begin{equation}\label{Lagrangianlorentz}
{\mathcal L}=\frac{1}{2}(\dot{A_i}-\nabla_i A_0)(\dot{A^i}-\nabla^i A_0) - \frac{1}{4}F_{ij}F^{ij}-\frac{1}{2}m^2A^\mu A_\mu-\frac{1}{2}X^{\mu\nu}A_\mu A_\nu\;.
\end{equation}
We have separated off the terms that involve time derivatives to identify the canonical momentum which is given by
\begin{equation}
\pi^i=\frac{\partial{\mathcal L}}{\partial\dot{A_i}}= \dot{A^i}-\nabla^i A_0\;.\label{momentum}
\end{equation}
The canonical Hamiltonian density is computed in the usual way to be
\begin{eqnarray}
{\mathcal H}&=&\pi^i\dot{A_i}-{\mathcal L}\label{Hamiltonian1}\\
&=&\frac{1}{2}\pi^i\pi_i-A_0\nabla_i\pi^i + \frac{1}{4}F_{ij}F^{ij}+\frac{1}{2}m^2A^\mu A_\mu+\frac{1}{2}X^{\mu\nu}A_\mu A_\nu\;,\label{Hamiltonian2}
\end{eqnarray}
where we have discarded a total derivative term to remove the differentiation from $A_0$ in the second term of \eqref{Hamiltonian2}. 

In the Faddeev-Jackiw method~\citep{FaddeevJackiw,Jackiwnotears} we regard the Lagrangian as a function of the variables $(\pi^i,A_i,\dot{A_i},A_0)$ which is first order in time derivatives and is found by inverting \eqref{Hamiltonian1} and \eqref{Hamiltonian2}. Because there is no dependence on $\dot{A_0}$ the associated field equation is simply the constraint
\begin{equation}
0=\frac{\partial{\mathcal L}}{\partial A_0}=\nabla_i\pi^i+m^2\,A_0-X^{00}\,A_0-X^{0i}\,A_i\;. \label{constraint}
\end{equation}
Assuming that $m^2\ne X^{00}$, this constraint can be solved for $A_0$ and the result can be substituted back into ${\mathcal L}$ to obtain the reduced Lagrangian that depends on the reduced set of variables $(\pi^i,A_i,\dot{A_i})$. The symplectic part of the Lagrangian is not affected by this reduction and retains the canonical form $\pi^i\dot{A_i}$. As noted by Faddeev and Jackiw~\cite{FaddeevJackiw,Jackiwnotears} this means that $\pi^i$ and $A_i$ are canonically conjugate variables with the usual commutation relations (or Poisson brackets in the classical theory) holding. In terms of the path integral formulation that we will use here, this means that the functional measure is formally simply an integral over $\pi^i$ and $A_i$ without any complicated measure factors occurring. The reduced Lagrangian density is
\begin{equation}\label{Lagrangianred}
{\mathcal L}_{\rm red}=\pi^i\dot{A_i} -\frac{1}{2}\pi^i\pi_i - \frac{1}{4}F_{ij}F^{ij}-\frac{1}{2}m^2A^i A_i-\frac{1}{2}X^{ij}A_i A_j-\frac{1}{2}(m^2-X^{00})^{-1}\big(\nabla_i\pi^i-X^{0i}A_i\big)^2\;.
\end{equation}
The partition function is then given by the functional integral expression
\begin{equation}
Z=\int\lbrack dA_i\rbrack\lbrack d\pi^i\rbrack\,e^{i\int dv_x{\mathcal L}_{\rm red}}\;,\label{Z}
\end{equation}
where we have denoted the invariant spacetime volume element by $dv_x=dtd^{N-1}x\,(\det g_{ij})^{1/2}$. This result agrees with the more general procedure described in \citep{TomsFJPImethod} in this case.

Because the reduced Lagrangian density \eqref{Lagrangianred} is quadratic in both $\pi^i$ and $A_i$ there is no impediment to computing the integration over both sets of variables. However to simplify the calculations we will specialize to the case where $X^{00}=0$ and $X^{0i}=0$ so that only the spatial components of $X^{\mu\nu}$ are non-zero. This means that the form of $X^{\mu\nu}$ is analogous to the assumption that the metric is ultrastatic in \eqref{ultrastatic}. It is possible to proceed without making these simplifying assumptions but this makes the calculations more unwieldy than necessary for the points that we are trying to make here. With these simplifications we can write \eqref{Lagrangianred} as
\begin{equation}\label{Lagrangianred2}
{\mathcal L}_{\rm red}=\pi^i\dot{A_i} -\frac{1}{2}\pi^i\Sigma_{ij}\pi^j - \frac{1}{4}F_{ij}F^{ij}-\frac{1}{2}m^2A^i A_i-\frac{1}{2}X^{ij}A_i A_j\;,
\end{equation}
where we have defined
\begin{equation}
\Sigma_{ij}=g_{ij}-\frac{1}{m^2}\,\nabla_i\nabla_j\;.\label{sigma}
\end{equation}
Note that $\Sigma_{ij}$ may be recognized as the spatial components of the operator that occurs in \eqref{BVidentity} which gives an explanation of why the identity described earlier works from the viewpoint of a constrained system. By translating the variable of momentum in the functional integral and computing the Gaussian integral it follows that
\begin{equation}
\int\lbrack d\pi^i\rbrack\,\exp\left\lbrace i\int dv_x(\pi^i\dot{A_i}-\frac{1}{2}\pi^i\Sigma_{ij}\pi^j)\right\rbrace =\left\lbrack \det\Sigma_{ij}\right\rbrack^{-1/2} \,\exp\left\lbrace \frac{i}{2}\int dv_x\dot{A_i}\widetilde{\Sigma}^{ij}\dot{A_j}\right\rbrace
\end{equation}
where we have defined $\widetilde{\Sigma}^{ij}$ to be the inverse of $\Sigma_{ij}$ given in \eqref{sigma}:
\begin{equation}
\Sigma_{ij}\widetilde{\Sigma}^{jk}=\delta^{k}_{i}\;.\label{sigmatilde}
\end{equation}
The partition function in \eqref{Z} becomes
\begin{equation}
Z=\left\lbrack \det\Sigma_{ij}\right\rbrack^{-1/2} \,\int\lbrack dA_i\rbrack\,e^{i\int dv_x\tilde{\mathcal L}_{\rm red}}\;,\label{Z2}
\end{equation}
where $\tilde{\mathcal L}_{\rm red}$ is
\begin{equation}\label{Lagrangianred3}
\tilde{\mathcal L}_{\rm red}=  - \frac{1}{4}F_{ij}F^{ij}-\frac{1}{2}m^2A^i A_i-\frac{1}{2}X^{ij}A_i A_j +\frac{1}{2}\dot{A_i}\widetilde{\Sigma}^{ij}\dot{A_j}\;.
\end{equation}

At this stage we will perfom the Wick rotation of the time coordinate to recover the Riemannian metric used earlier in Sec.~\ref{Proca} in order to compare results. In place of $Z$ we will use the one-loop effective action which is
\begin{equation}
\Gamma_{\rm 1-loop}=\frac{1}{2}\ln\det\Sigma_{ij}-\ln\int\lbrack dA_i\rbrack\,e^{-S_E}\;,\label{gamma 1 loop}
\end{equation}
with 
\begin{equation}
S_E=\int dv_x\left(\frac{1}{4}F_{ij}F^{ij}+\frac{1}{2}m^2A^i A_i+\frac{1}{2}X^{ij}A_i A_j +\frac{1}{2}\dot{A_i}\,\widetilde{\Sigma}^{ij}\dot{A_j} \right).\label{SE1}
\end{equation}
After integrating by parts we can write the action $S_E$ in the generic form
\begin{equation}
S_E=\frac{1}{2}\int dv_x\,A_i\,\Delta^{ij}\, A_j\;,\label{SE}
\end{equation}
with the operator $\Delta^{ij}$ in this case defined by
\begin{equation}
\Delta^{ij}= -g^{ij}\nabla^2+\nabla^j\nabla^i+m^2\,g^{ij}+X^{ij}-\widetilde{\Sigma}^{ij}\partial^2_0.\label{deltaij}
\end{equation}
The remaining functional integration in \eqref{gamma 1 loop} can then be computed with the result that
\begin{eqnarray}
\Gamma_{\rm 1-loop}&=&\frac{1}{2}\ln\det\Sigma_{ij}+\frac{1}{2}\ln\det\Delta^{ij}\nonumber\\
&=&\frac{1}{2}\ln\det(\Sigma_{ij}\Delta^{jk})\;,\label{gamma1result}
\end{eqnarray}
if we combine the two terms. Because the free indices inside of the $\ln\det$ are mixed the result here is invariant under any coordinate transformation that preserves the ultrastatic form of the metric. This is consistent with the claim that the measure used to compute the functional integral is trivial without the need for any measure factors that involve the metric for example. (See \cite{Tomsfunctionalmeasure} for example.) After using \eqref{sigma} and \eqref{sigmatilde} along with some curvature identities it follows that
\begin{eqnarray}
\Sigma_{ij}\Delta^{jk}&=&\delta_{i}^{k}(-\Box+m^2)+R_{i}{}^{k}+X_{i}{}^{k}\nonumber\\
&&\quad -\frac{1}{m^2}\left\lbrack \nabla_i\nabla_j X^{jk}+\nabla_jX^{jk}\nabla_i+\nabla_iX^{jk}\nabla_j+X^{jk}\nabla_i\nabla_j\right\rbrack.\label{deltatilde}
\end{eqnarray}
In the absence of the non-minimal term $X^{ij}$ the right hand side of this expression can be observed to coincide with the spatial components of the result in \eqref{BVidentity}. It should now be readily apparent that the operator required for the computation of the one-loop effective action is non-minimal in the sense of Barvinsky and Vilkovisky~\citep{BarvinskyVilkovisky} if $X^{ij}\ne0$. The expression defined earlier in \eqref{oneloop} is not the effective action in the case where $X^{\mu\nu}\ne0$ in general.

\subsection{One-loop effective action in an ultrastatic spacetime}\label{sec-ultra}

If we set $X^{ij}=0$, but keep the general ultrastatic metric, then \eqref{deltatilde} becomes
\begin{equation}
\Sigma_{ij}\Delta^{jk}=\delta_{i}^{k}(-\Box+m^2)+R_{i}{}^{k}.\label{B1}
\end{equation}
This is just the minimal vector operator but with just the spatial components occurring. The local momentum space method can be used as described earlier to find
\begin{eqnarray}
(G_0)^{i}{}_{j}&=&\delta^{i}_{j}\,S,\label{B2}\\
(G_2)^{i}{}_{j}&=&\delta^{i}_{j}\left(\frac{1}{3}\,R\,S^2-\frac{2}{3}\,R_{kl}\,p^kp^l\,S^3\right)-\frac{2}{3}\,R^{i}{}_{j}\,S^2-\frac{4}{3}\,R^{i}{}_{kjl}p^kp^l\,S^3,\label{B3}\\
(G_4)^{i}{}_{j}&=&\delta^{i}_{j}\,S^3\left( \frac{1}{9}\,R^2+\frac{2}{45}\,R_{kl}R^{kl}+\frac{1}{15}\,R_{klmn}R^{klmn}+\frac{2}{5}\,\Box R\right)\nonumber\\
&&+S^3\left( R^{ik}R_{jk}-\frac{2}{3}\,RR^{i}{}_{j}-\frac{1}{2}\,R^{iklm}R_{jklm} -\Box R^{i}{}_{j}\right)\nonumber\\
&&+\delta^{i}_{j}\,S^4\,p^kp^l\left(\frac{4}{5}R_{mn}R^{mknl}-\frac{2}{3}RR_{kl}-\frac{8}{5}R_{km}R^{m}{}_{l} -\frac{8}{15}R_{pmnk}R^{pmn}{}_{l}\right.\nonumber\\
&&\qquad\qquad\left.-\frac{12}{5}R_{;kl}-\frac{4}{5}\Box R_{kl} \right)\nonumber\\
&&+\delta^{i}_{j}\,S^5\,p^kp^lp^mp^n\left(\frac{4}{3}R_{kl}R_{mn}+\frac{16}{5}R_{pkql}R^{p}{}_{m}{}^{q}{}_{n}+\frac{24}{5}R_{kl;mn} \right)\nonumber\\
&&+2\,S^4\,p^kp^l\left(R^{imn}{}_{k}R_{jmnl}+R^{i}{}_{j}R_{kl}+ R_{jk}{}^{;i}{}_{l} - R^{i}{}_{k;jl}+2\,R^{i}{}_{j;kl}\right).\label{B4}
\end{eqnarray}

We can now use \eqref{gammmaderiv} but where only the spatial indices are summed over as is relevant from \eqref{B1}. The pole part of the one-loop effective action is then contained in (for $N=4$)
\begin{equation}\label{B5}
\frac{\partial}{\partial m^2}\Gamma_{\rm 1-loop}=\frac{1}{2}\int dv_x\int\frac{d^Np}{(2\pi)^N}\Big\lbrack (G_0)^{i}{}_{i}+ (G_2)^{i}{}_{i} + (G_4)^{i}{}_{i} + \cdots\Big\rbrack.
\end{equation}
It is straightforward to use \eqref{B2}--\eqref{B4} and the momentum integration in \eqref{momint} and \eqref{Ik} to find
\begin{eqnarray}
\Gamma_{\rm 1-loop}&=&(4\pi)^{-N/2}\int dv_x \Big\lbrace -\frac{(N-1)}{2}\,\Gamma(-N/2) \, (m^2)^{N/2}\nonumber\\
&&\qquad+\frac{(7-N)}{12}\,\Gamma(1-N/2)\, (m^2)^{N/2-1}\nonumber\\
&&\qquad+\Gamma(2-N/2)\,(m^2)^{N/2-2}\Big\lbrack \frac{(5-N)}{60}\,\Box R + \frac{(N-91)}{360}\,R_{\mu\nu}R^{\mu\nu}\nonumber\\
&&\qquad\qquad + \frac{(16-N)}{360}\,R_{\mu\nu\lambda\sigma}R^{\mu\nu\lambda\sigma}+\frac{(13-N)}{144}\,R^2\Big\rbrack\Big\rbrace .\label{B6}
\end{eqnarray}
It is not necessary to apply the $\delta(0)\rightarrow0$ rule here as it was before as there are now no quartic divergences present. We have written the result back in terms of the full curvature tensors here since this must follow on the basis of general covariance. We have included the $\Box R$ term here even though it is a total derivative just to demonstrate that it can be calculated without difficulty here as mentioned earlier where the previous calculation ran into some difficulty. It can also be noted that the coefficients of the curvature squared terms are consistent with those that we found in the evaluation of the heat kernel coefficient in \eqref{HK7}. It is now a simple calculation to expand about the pole at $N=4$. If this is done then the pole part of \eqref{B6} agrees precisely with what we found earlier in \eqref{onelooppole} with the added bonus of the $\Box R$ term.

\subsection{Generalized Proca model in flat spacetime and renormalizability}\label{sec-flat}

We will now specialize to flat spacetime with the simplifying assumption that $X^{ij}$ is constant. This will remove all but the last term in \eqref{deltatilde} to leave
\begin{equation}
\Sigma_{ij}\Delta^{jk}=\delta_{i}^{k}(-\Box+m^2)+X_{i}{}^{k}-\frac{1}{m^2}\,X^{jk}\partial_i\partial_j.\label{C1}
\end{equation}
The one-loop effective action is given by \eqref{gamma1result}. An appealing approach would be to treat $X_{i}{}^{k}$ perturbatively; however, this proves problematic for computing even the pole part of the effective action since it turns out that pole terms come from all powers of $X^{ij}$. If one proceeds to do this calculation regardless then the terms up to second order agree precisely with what was found from the heat kernel coefficients in sec.~\ref{sec-heat} specialized to this case; however, the full pole terms of the effective action are not found. The operator that we are dealing with in \eqref{C1} is a generalization of one dealt with in \cite{MossToms} where it was found that the dependence on a non-minimal parameter for a simpler operator was not analytic in general. This non-analyticity would be expected to carry over for the more complicated operator \eqref{C1} and we will show that it does and explains why a naive perturbative approach fails. 

The case of arbitrary $X^{ij}$, even for the restricted case of a constant expression, is still intractable. We will therefore limit the discussion to two very simple cases: $X^{ij}=X\delta^{ij}$ where $X$ is constant, and $X^{ij}=x^ix^j$ where $x^i$ is a constant vector. In both cases we will find an exact result for the one-loop effective action and show that it is not renormalizable by local counterterms. The reason why these cases are both starightforward is that there is a simple relation between arbitrary powers of $X^{ij}$ and a single power of $X^{ij}$ that is not true for general $X^{ij}$.

\subsubsection{$X^{ij}=X\delta^{ij}$}\label{secSdelta}

With $X^{ij}=X\delta^{ij}$ assumed the operator in \eqref{C1} results in
\begin{equation}
\Gamma_{\rm 1-loop}=\frac{1}{2}\ln\det\lbrack\delta_{i}^{k}(-\Box+m^2+X)-\frac{1}{m^2}\,X\,\partial_i\partial^k\rbrack.\label{C2}
\end{equation}
One way to work out \eqref{C2} is to evaluate the eigenvalues for the operator that occurs. Define $\psi_k$ to be the eigenfunction that obeys
\begin{equation}
\lbrack\delta_{i}^{k}(-\Box+m^2+X)-\frac{1}{m^2}\,X\,\partial_i\partial^k\rbrack\,\psi_k=\lambda\,\psi_i.\label{C3}
\end{equation}
Decompose $\psi_k$ into a transverse part $\psi^\perp_k$ that obeys
\begin{equation}
\partial^k\psi^\perp_k=0,\label{C4}
\end{equation}
and a longitudinal part $\partial_k\psi$:
\begin{equation}
\psi_k=\psi^\perp_k+\partial_k\psi.\label{C5}
\end{equation}
It is then easy to see from \eqref{C3} that 
\begin{eqnarray}
&&(-\Box+m^2+X)\,\psi^\perp_k=\lambda\,\psi^\perp_k,\label{C6}\\
&&\Big\lbrack -\partial^2_0-\Big(1+\frac{X}{m^2}\Big)\nabla^2+m^2+X\Big\rbrack\,\psi=\lambda\,\psi.\label{C7}
\end{eqnarray}
Note that $\Box=\partial^2_0+\nabla^2$ with $\nabla^2$ the spatial Laplacian. The eigenvalue for $\psi^\perp_k$ in \eqref{C6} is $(N-2)$-fold degenerate whereas that for $\psi$ in \eqref{C7} is not degenerate. This establishes that
\begin{equation}
\Gamma_{\rm 1-loop}=\frac{1}{2}(N-2)\ln\det(-\Box+m^2+X) + \frac{1}{2}\ln\det\lbrack -\partial^2_0-\Big(1+\frac{X}{m^2}\Big)\nabla^2+m^2+X\Big\rbrack.\label{C8}
\end{equation}

Before analyzing the result in \eqref{C8} we will verify it using another method that expands the logarithm of \eqref{C2} as in the derivation of \eqref{BVidentity} in \citep{BarvinskyVilkovisky}. Split the operator in \eqref{C3} into two terms:
\begin{eqnarray}
(\Delta_0)_{i}{}^{j}&=&\delta_{i}^{j}(-\Box+m^2+X),\label{C9}\\
D_{i}{}^{j}&=&-\frac{1}{m^2}\,X\,\partial_i\partial^j.\label{C10}
\end{eqnarray}
From \eqref{C2}, making use of the identity $\ln\det={\rm tr}\ln$ we have
\begin{eqnarray}
\Gamma_{\rm 1-loop}&=&\frac{1}{2}{\rm tr}\ln(\Delta_0+D)\nonumber\\
&=&\frac{1}{2}{\rm tr}\ln\,\Delta_0+ \frac{1}{2}{\rm tr}\ln(I+\Delta_0^{-1}D).\label{C11}
\end{eqnarray}
Here $\Delta_0^{-1}$ is just the inverse of \eqref{C9} which is of course just the Green's function for the operator. The logarithm in the second term of \eqref{C11} can now be expanded resulting in
\begin{equation}
\Gamma_{\rm 1-loop}=\frac{1}{2}{\rm tr}\ln\,\Delta_0-\frac{1}{2}\sum_{n=1}^{\infty} \frac{(-1)^n}{n}\,{\rm tr}\lbrack (\Delta_0^{-1}D)^n\rbrack.\label{C12}
\end{equation}
Making use of \eqref{C9} and \eqref{C10} shows that
\begin{equation}
\lbrack(\Delta_0^{-1}D)^n\rbrack_{i}{}^{j}=\left(-\frac{X}{m^2}\right)^n(-\Box+m^2+X)^{-n}(\nabla^2)^{n-1}\partial_i\partial^j.\label{C13}
\end{equation}
When \eqref{C13} is used back in \eqref{C12} the sum can be performed in terms of a logarithm and the result of \eqref{C8} is regained.

Having now established \eqref{C8} by two different methods we turn to its evaluation. The $\ln\det={\rm tr}\ln$ identity can be used again and the trace evaluated using a momentum space basis. The one-loop effective Lagrangian just differs from the one-loop effective action by a spacetime volume integration. It then follows that 
\begin{eqnarray}
{\mathcal L}_{\rm 1-loop}&=&\frac{1}{2}(N-2)\int\frac{d^Np}{(2\pi)^N}\,\ln(p^2+m^2+X)\nonumber\\
&&\quad+\frac{1}{2} \int\frac{d^Np}{(2\pi)^N}\,\ln\lbrack p_0^2+m^2+X+(1+X/m^2){\mathbf p}^2\rbrack. \label{C14}
\end{eqnarray} 
Here $p^2=p_0^2+{\mathbf p}^2$ where ${\mathbf p}$ denotes the spatial components of the momentum. We can make use of the Schwinger proper time representation~\cite{SchwingerQED2,Schwingerproper} to find 
\begin{eqnarray}
{\mathcal L}_{\rm 1-loop}&=&\frac{1}{2}(N-2)\int\frac{d^Np}{(2\pi)^N}\,\int\limits_{0}^{\infty}\frac{d\tau}{\tau}\,e^{-\tau(p^2+m^2+X)}\nonumber\\
&&+\frac{1}{2}\int\frac{d^Np}{(2\pi)^N}\,\int\limits_{0}^{\infty}\frac{d\tau}{\tau}\,e^{-\tau\lbrack p_0^2+m^2+X+(1+X/m^2){\mathbf p}^2\rbrack}. \label{C15}
\end{eqnarray}
The momentum integration is easily performed and the integration over $\tau$ done using the standard integral representation for the $\Gamma$-function. The end result is
\begin{equation}
{\mathcal L}_{\rm 1-loop}=\frac{1}{2}(4\pi)^{-N/2}\Gamma(-N/2)\left\lbrack (N-2)(m^2+X)^{N/2} + (m^2+X)^{1/2}(m^2)^{(N-1)/2}\right\rbrack.\label{C16}
\end{equation}
As a check on this result, if we let $X\rightarrow0$ so that we have the minimal Proca model, we obtain $(N-1)$ times the real scalar field result as would be expected. 

The presence of the factor of $(m^2+X)^{1/2}$ in the second term of \eqref{C16} makes the theory non-renormalizable with the addition of local Lagrangian counterterms. If we expand about $N=4$ for example we find
\begin{equation}
{\mathcal L}_{\rm 1-loop}=-\frac{1}{32\pi^2(N-4)}\lbrack 2(m^2+X)^2+m^3(m^2+X)^{1/2}\rbrack+\cdots.\label{C17}
\end{equation}
Although the first term of \eqref{C17} could be absorbed by a renormalization of the cosmological constant, $X^{ij}$ and a term quadratic in $X^{ij}$ there is no way to absorb the divergence of the second term in such a way. It is also clear from this result why an expansion in powers of $X$ will not obtain the correct pole term unless the expansion is carried out to all orders; the first few terms will not suffice. The origin of this non-renormalizability resides in the non-minimal nature of the operator involved.

\subsubsection{$X^{ij}=x^ix^j$}

We will make use of momentum space from the start in this section and write from \eqref{C2}
\begin{equation}
{\mathcal L}_{\rm 1-loop}=\frac{1}{2}\int\frac{d^Np}{(2\pi)^N}\,{\rm tr}\ln\lbrack (\tilde{\Delta}_0)_{i}{}^{j}+y_i x^j\rbrack,\label{C18}
\end{equation}
where 
\begin{eqnarray}
(\tilde{\Delta}_0)_{i}{}^{j}&=&\delta_{i}{}^{j}(p^2+m^2),\label{C19}\\
y_i&=&x_i+\frac{1}{m^2}({\mathbf p}\cdot{\mathbf x})\,p_i.\label{C20}
\end{eqnarray}
Again $p^2=p_0^2+{\mathbf p}^2$ and here ${\mathbf p}\cdot{\mathbf x}=p_ix^i$ involves just the spatial components. The logarithm can be expanded as in \eqref{C11} and \eqref{C12} and this time it can be noted that
\begin{equation}
{\rm tr}\ln\lbrack\delta_{i}{}^{j}+(\tilde{\Delta}_0^{-1})_{i}{}^{k}\,y_kx^j\rbrack=\ln\left\lbrack 1+ \frac{{\mathbf x}\cdot{\mathbf y}}{p^2+m^2}\right\rbrack.\label{C21}
\end{equation}
We then find from \eqref{C18} that
\begin{eqnarray}
{\mathcal L}_{\rm 1-loop}&=&\frac{1}{2}(N-2)\int\frac{d^Np}{(2\pi)^N}\,\ln(p^2+m^2)\nonumber\\
&&+\frac{1}{2}\int\frac{d^Np}{(2\pi)^N}\,\ln\left\lbrack p^2+m^2+x^2+\frac{1}{m^2} ({\mathbf p}\cdot{\mathbf x})^2\right\rbrack,\label{C22}
\end{eqnarray}
where $x^2=x_ix^i$. The presence of the nonminimal operator is observed in the $({\mathbf p}\cdot{\mathbf x})^2$ term that occurs. We can again make use of the Schwinger proper time representation~\cite{SchwingerQED2,Schwingerproper} and find
\begin{equation}
{\mathcal L}_{\rm 1-loop}=\frac{1}{2}(4\pi)^{-N/2}\Gamma(-N/2)\left\lbrack (N-2)(m^2)^{N/2}+(m^2)^{1/2}(m^2+x^2)^{(N-1)/2}\right\rbrack.\label{C23}
\end{equation}
As in Sec.~\ref{secSdelta} the second term coming from the non-minimal operator can give rise to a divergence that is not local in $X^{ij}$. This occurs for even spacetime dimensions. If we take $N\rightarrow4$ and note that $x^2={\rm tr}X$ here we have
\begin{equation}
{\mathcal L}_{\rm 1-loop}=-\frac{1}{32\pi^2(N-4)}\lbrack 2m^4+m(m^2+{\rm tr}X)^{3/2}\rbrack+\cdots.\label{C24}
\end{equation}
As before if we let $X^{ij}\rightarrow0$ we recover the real scalar field result mutiplied by the factor $(N-1)$.

\section{Discussion}

We have discussed the quantization of the massive vector (Proca) field with and without non-minimal terms in both flat and curved spacetime. In particular we used Faddeev-Jackiw~\citep{FaddeevJackiw} quantization to calculate the one-loop divergences in the effective action in both flat and curved spacetimes. Some subtleties of the quantization procedure and the use of heat kernel methods were considered. We showed that the addition of non-minimal terms to the theory will in general result in a theory that is not renormalizable with the addition of local counterterms to the Lagrangian.

It would be interesting to consider the case of a self-interacting Proca field in light of the results found here. Because the gauge invariance is lost by the presence of a mass or the non-minimal terms it is possible to add on a self-interaction of the form $(A_\mu A^\mu)^2$ for example that is renormalizable by power counting. It would be interesting to see if the one-loop divergences in this case are local polynomials. The Faddeev-Jackiw~\citep{FaddeevJackiw} approach leads to a clear path integral expression. It would also be interesting to examine the theory from the Steuckelberg approach. We hope to report on this elsewhere.


\begin{thebibliography}{33}
\expandafter\ifx\csname natexlab\endcsname\relax\def\natexlab#1{#1}\fi
\expandafter\ifx\csname bibnamefont\endcsname\relax
  \def\bibnamefont#1{#1}\fi
\expandafter\ifx\csname bibfnamefont\endcsname\relax
  \def\bibfnamefont#1{#1}\fi
\expandafter\ifx\csname citenamefont\endcsname\relax
  \def\citenamefont#1{#1}\fi
\expandafter\ifx\csname url\endcsname\relax
  \def\url#1{\texttt{#1}}\fi
\expandafter\ifx\csname urlprefix\endcsname\relax\def\urlprefix{URL }\fi
\providecommand{\bibinfo}[2]{#2}
\providecommand{\eprint}[2][]{\url{#2}}

\bibitem[{\citenamefont{Higgs}(1964{\natexlab{a}})}]{higgs1964broken1}
\bibinfo{author}{\bibfnamefont{P.~W.} \bibnamefont{Higgs}},
  \bibinfo{journal}{Phys. Lett.} \textbf{\bibinfo{volume}{12}},
  \bibinfo{pages}{132} (\bibinfo{year}{1964}{\natexlab{a}}).

\bibitem[{\citenamefont{Higgs}(1964{\natexlab{b}})}]{higgs1964broken}
\bibinfo{author}{\bibfnamefont{P.~W.} \bibnamefont{Higgs}},
  \bibinfo{journal}{Phys. Rev. Lett.} \textbf{\bibinfo{volume}{13}},
  \bibinfo{pages}{508} (\bibinfo{year}{1964}{\natexlab{b}}).

\bibitem[{\citenamefont{Englert and Brout}(1964)}]{englert1964broken}
\bibinfo{author}{\bibfnamefont{F.}~\bibnamefont{Englert}} \bibnamefont{and}
  \bibinfo{author}{\bibfnamefont{R.}~\bibnamefont{Brout}},
  \bibinfo{journal}{Phys. Rev. Lett.} \textbf{\bibinfo{volume}{13}},
  \bibinfo{pages}{321} (\bibinfo{year}{1964}).

\bibitem[{\citenamefont{Guralnik et~al.}(1964)\citenamefont{Guralnik, Hagen,
  and Kibble}}]{guralnik1964global}
\bibinfo{author}{\bibfnamefont{G.~S.} \bibnamefont{Guralnik}},
  \bibinfo{author}{\bibfnamefont{C.~R.} \bibnamefont{Hagen}}, \bibnamefont{and}
  \bibinfo{author}{\bibfnamefont{T.~W.} \bibnamefont{Kibble}},
  \bibinfo{journal}{Phys. Rev. Lett.} \textbf{\bibinfo{volume}{13}},
  \bibinfo{pages}{585} (\bibinfo{year}{1964}).

\bibitem[{\citenamefont{Proca}(1936)}]{proca1936theorie}
\bibinfo{author}{\bibfnamefont{A.~L.} \bibnamefont{Proca}},
  \bibinfo{journal}{J. Phys. Radium} \textbf{\bibinfo{volume}{7}},
  \bibinfo{pages}{347} (\bibinfo{year}{1936}).

\bibitem[{\citenamefont{Pauli}(1941)}]{pauli1941relativistic}
\bibinfo{author}{\bibfnamefont{W.}~\bibnamefont{Pauli}}, \bibinfo{journal}{Rev.
  Mod. Phys.} \textbf{\bibinfo{volume}{13}}, \bibinfo{pages}{203}
  (\bibinfo{year}{1941}).

\bibitem[{\citenamefont{Wentzel}(1949)}]{Wentzel}
\bibinfo{author}{\bibfnamefont{G.}~\bibnamefont{Wentzel}},
  \emph{\bibinfo{title}{Quantum Theory of Fields}}
  (\bibinfo{publisher}{Interscience}, \bibinfo{year}{1949}).

\bibitem[{\citenamefont{Ruegg and Ruiz-Altaba}(2004)}]{ruegg2004stueckelberg}
\bibinfo{author}{\bibfnamefont{H.}~\bibnamefont{Ruegg}} \bibnamefont{and}
  \bibinfo{author}{\bibfnamefont{M.}~\bibnamefont{Ruiz-Altaba}},
  \bibinfo{journal}{Int. J. Mod. Phys. A} \textbf{\bibinfo{volume}{19}},
  \bibinfo{pages}{3265} (\bibinfo{year}{2004}).

\bibitem[{\citenamefont{Furlani}(1997)}]{furlani1997quantization}
\bibinfo{author}{\bibfnamefont{E.~P.} \bibnamefont{Furlani}},
  \bibinfo{journal}{Class. and Quantum Grav.} \textbf{\bibinfo{volume}{14}},
  \bibinfo{pages}{1665} (\bibinfo{year}{1997}).

\bibitem[{\citenamefont{Barvinsky and Vilkovisky}(1985)}]{BarvinskyVilkovisky}
\bibinfo{author}{\bibfnamefont{A.~O.} \bibnamefont{Barvinsky}}
  \bibnamefont{and} \bibinfo{author}{\bibfnamefont{G.~A.}
  \bibnamefont{Vilkovisky}}, \bibinfo{journal}{Physics Reports}
  \textbf{\bibinfo{volume}{119}}, \bibinfo{pages}{1} (\bibinfo{year}{1985}).

\bibitem[{\citenamefont{Gorbar and Shapiro}(2003)}]{gorbar2003renormalization}
\bibinfo{author}{\bibfnamefont{E.~V.} \bibnamefont{Gorbar}} \bibnamefont{and}
  \bibinfo{author}{\bibfnamefont{I.~L.} \bibnamefont{Shapiro}},
  \bibinfo{journal}{JHEP} \textbf{\bibinfo{volume}{2003}}, \bibinfo{pages}{004}
  (\bibinfo{year}{2003}).

\bibitem[{\citenamefont{Buchbinder et~al.}(2007)\citenamefont{Buchbinder,
  de~Berredo-Peixoto, and Shapiro}}]{buchbinder2007quantum}
\bibinfo{author}{\bibfnamefont{I.}~\bibnamefont{Buchbinder}},
  \bibinfo{author}{\bibfnamefont{G.}~\bibnamefont{de~Berredo-Peixoto}},
  \bibnamefont{and} \bibinfo{author}{\bibfnamefont{I.}~\bibnamefont{Shapiro}},
  \bibinfo{journal}{Phys. Lett. B} \textbf{\bibinfo{volume}{649}},
  \bibinfo{pages}{454} (\bibinfo{year}{2007}).

\bibitem[{\citenamefont{Prescod-Weinstein and
  Bertschinger}(2015)}]{prescod2015extension}
\bibinfo{author}{\bibfnamefont{C.}~\bibnamefont{Prescod-Weinstein}}
  \bibnamefont{and}
  \bibinfo{author}{\bibfnamefont{E.}~\bibnamefont{Bertschinger}},
  \bibinfo{journal}{Class. and Quantum Grav.} \textbf{\bibinfo{volume}{32}},
  \bibinfo{pages}{075011} (\bibinfo{year}{2015}).

\bibitem[{\citenamefont{Faddeev and Jackiw}(1988)}]{FaddeevJackiw}
\bibinfo{author}{\bibfnamefont{L.}~\bibnamefont{Faddeev}} \bibnamefont{and}
  \bibinfo{author}{\bibfnamefont{R.}~\bibnamefont{Jackiw}},
  \bibinfo{journal}{Phys. Rev. Lett.} \textbf{\bibinfo{volume}{60}},
  \bibinfo{pages}{1692} (\bibinfo{year}{1988}).

\bibitem[{\citenamefont{Novello and Salim}(1979)}]{novello1979nonlinear}
\bibinfo{author}{\bibfnamefont{M.}~\bibnamefont{Novello}} \bibnamefont{and}
  \bibinfo{author}{\bibfnamefont{J.}~\bibnamefont{Salim}},
  \bibinfo{journal}{Phys. Rev. D} \textbf{\bibinfo{volume}{20}},
  \bibinfo{pages}{377} (\bibinfo{year}{1979}).

\bibitem[{\citenamefont{Davies and Toms}(1985)}]{davies1985boundary}
\bibinfo{author}{\bibfnamefont{P.~C.~W.} \bibnamefont{Davies}}
  \bibnamefont{and} \bibinfo{author}{\bibfnamefont{D.~J.} \bibnamefont{Toms}},
  \bibinfo{journal}{Phys. Rev. D} \textbf{\bibinfo{volume}{31}},
  \bibinfo{pages}{1363} (\bibinfo{year}{1985}).

\bibitem[{\citenamefont{Bunch and Parker}(1979)}]{BunchParker}
\bibinfo{author}{\bibfnamefont{T.~S.} \bibnamefont{Bunch}} \bibnamefont{and}
  \bibinfo{author}{\bibfnamefont{L.}~\bibnamefont{Parker}},
  \bibinfo{journal}{Phys. Rev. D} \textbf{\bibinfo{volume}{20}},
  \bibinfo{pages}{2499} (\bibinfo{year}{1979}).

\bibitem[{\citenamefont{Toms}(2014)}]{toms2014local}
\bibinfo{author}{\bibfnamefont{D.~J.} \bibnamefont{Toms}},
  \bibinfo{journal}{Phys. Rev. D} \textbf{\bibinfo{volume}{90}},
  \bibinfo{pages}{044072} (\bibinfo{year}{2014}).

\bibitem[{\citenamefont{Misner et~al.}(1973)\citenamefont{Misner, Thorne, and
  Wheeler}}]{MTW}
\bibinfo{author}{\bibfnamefont{C.~W.} \bibnamefont{Misner}},
  \bibinfo{author}{\bibfnamefont{K.~S.} \bibnamefont{Thorne}},
  \bibnamefont{and} \bibinfo{author}{\bibfnamefont{J.~A.}
  \bibnamefont{Wheeler}}, \emph{\bibinfo{title}{{Gravitation}}}
  (\bibinfo{publisher}{W. H. Freeman}, \bibinfo{year}{1973}).

\bibitem[{\citenamefont{Toms}(1982)}]{TomsPRDscalar}
\bibinfo{author}{\bibfnamefont{D.~J.} \bibnamefont{Toms}},
  \bibinfo{journal}{Phys. Rev. D} \textbf{\bibinfo{volume}{26}},
  \bibinfo{pages}{2713} (\bibinfo{year}{1982}).

\bibitem[{\citenamefont{DeWitt}(1965)}]{DeWittdynamical}
\bibinfo{author}{\bibfnamefont{B.~S.} \bibnamefont{DeWitt}},
  \emph{\bibinfo{title}{Dynamical Theory of Groups and Fields}}
  (\bibinfo{publisher}{Gordon and Breach}, \bibinfo{year}{1965}).

\bibitem[{\citenamefont{Moss and Toms}(2014)}]{MossToms}
\bibinfo{author}{\bibfnamefont{I.~G.} \bibnamefont{Moss}} \bibnamefont{and}
  \bibinfo{author}{\bibfnamefont{D.~J.} \bibnamefont{Toms}},
  \bibinfo{journal}{J. Phys. A: Math. and Theor.}
  \textbf{\bibinfo{volume}{47}}, \bibinfo{pages}{215401 (28pp)}
  (\bibinfo{year}{2014}).

\bibitem[{\citenamefont{Dirac}(1964)}]{DiracLecturesQM}
\bibinfo{author}{\bibfnamefont{P.~A.~M.} \bibnamefont{Dirac}},
  \emph{\bibinfo{title}{Lectures on Quantum Mechanics}}
  (\bibinfo{publisher}{Belfer Graduate School}, \bibinfo{year}{1964}).

\bibitem[{\citenamefont{Sundermeyer}(1982)}]{Sundermeyer}
\bibinfo{author}{\bibfnamefont{K.}~\bibnamefont{Sundermeyer}},
  \emph{\bibinfo{title}{Constrained Dynamics}}, vol. \bibinfo{volume}{169} of
  \emph{\bibinfo{series}{Lecture Notes in Physics}}
  (\bibinfo{publisher}{Springer-Verlag}, \bibinfo{year}{1982}).

\bibitem[{\citenamefont{Gitman and Tyutin}(1990)}]{GitmanTyutin}
\bibinfo{author}{\bibfnamefont{D.~M.} \bibnamefont{Gitman}} \bibnamefont{and}
  \bibinfo{author}{\bibfnamefont{I.~V.} \bibnamefont{Tyutin}},
  \emph{\bibinfo{title}{Quantization of Fields with Constraints}}
  (\bibinfo{publisher}{Springer-Verlag}, \bibinfo{year}{1990}).

\bibitem[{\citenamefont{Henneaux and Teitelboim}(1991)}]{HenneauxTeitelboim}
\bibinfo{author}{\bibfnamefont{M.}~\bibnamefont{Henneaux}} \bibnamefont{and}
  \bibinfo{author}{\bibfnamefont{C.}~\bibnamefont{Teitelboim}},
  \emph{\bibinfo{title}{Quantization of Gauge Systems}}
  (\bibinfo{publisher}{Princeton University Press}, \bibinfo{year}{1991}).

\bibitem[{\citenamefont{Rothe and Rothe}(2010)}]{RotheandRothe}
\bibinfo{author}{\bibfnamefont{H.~J.} \bibnamefont{Rothe}} \bibnamefont{and}
  \bibinfo{author}{\bibfnamefont{K.~D.} \bibnamefont{Rothe}},
  \emph{\bibinfo{title}{Classical and Quantum Dynamics of Constrained
  Hamiltonian Systems}}, vol.~\bibinfo{volume}{81} of
  \emph{\bibinfo{series}{Lecture Notes in Physics}} (\bibinfo{publisher}{World
  Scientific}, \bibinfo{year}{2010}).

\bibitem[{\citenamefont{Jackiw}(1993)}]{Jackiwnotears}
\bibinfo{author}{\bibfnamefont{R.}~\bibnamefont{Jackiw}},
  \bibinfo{journal}{arXiv:hep-th/930607v1}  (\bibinfo{year}{1993}).

\bibitem[{\citenamefont{Fulling}(1989)}]{fulling1989aspects}
\bibinfo{author}{\bibfnamefont{S.~A.} \bibnamefont{Fulling}},
  \emph{\bibinfo{title}{Aspects of quantum field theory in curved spacetime}}
  (\bibinfo{publisher}{Cambridge University Press}, \bibinfo{year}{1989}).

\bibitem[{\citenamefont{Toms}(2015)}]{TomsFJPImethod}
\bibinfo{author}{\bibfnamefont{D.~J.} \bibnamefont{Toms}},
  \bibinfo{journal}{arXiv:1508.07432 [hep-th]}  (\bibinfo{year}{2015}).

\bibitem[{\citenamefont{Toms}(1987)}]{Tomsfunctionalmeasure}
\bibinfo{author}{\bibfnamefont{D.~J.} \bibnamefont{Toms}},
  \bibinfo{journal}{Phys. Rev. D} \textbf{\bibinfo{volume}{35}},
  \bibinfo{pages}{3796} (\bibinfo{year}{1987}).

\bibitem[{\citenamefont{Schwinger}(1949)}]{SchwingerQED2}
\bibinfo{author}{\bibfnamefont{J.}~\bibnamefont{Schwinger}},
  \bibinfo{journal}{Phys. Rev.} \textbf{\bibinfo{volume}{75}},
  \bibinfo{pages}{651} (\bibinfo{year}{1949}).

\bibitem[{\citenamefont{Schwinger}(1951)}]{Schwingerproper}
\bibinfo{author}{\bibfnamefont{J.}~\bibnamefont{Schwinger}},
  \bibinfo{journal}{Phys. Rev.} \textbf{\bibinfo{volume}{82}},
  \bibinfo{pages}{664} (\bibinfo{year}{1951}).

\end{thebibliography}

\end{document}